\documentclass[
aps,%
12pt,%
final,%
notitlepage,%
oneside,%
onecolumn,%
nobibnotes,%
nofootinbib,%
superscriptaddress,%
showpacs,%
centertags]%
{revtex4}

\def\ls{\left(}
\def\rs{\right)}

\begin{document}

\title{Neutrino Oscillations in Gravitational Field.}

\author{\firstname{S.~I.}~\surname{Godunov}}
\email{sgodunov@itep.ru}
\affiliation{%
Moscow Institute for Physics and Technology, Russia
}%
\affiliation{%
Institute of Theoretical and Experimental Physics, Russia\\~
}%

\author{\firstname{G.~S.}~\surname{Pastukhov}}
\email{grigoriypas@gmail.com}
\affiliation{%
Moscow Institute for Physics and Technology, Russia
}%
\affiliation{%
Institute of Theoretical and Experimental Physics, Russia\\~
}%

\begin{abstract}
 We calculate the gravitational correction to the phase difference between
 neutrino mass eigenstates. There is a number of papers, where this correction was 
 calculated for the spherically symmetric gravitational field described by the Schwarzschild metric.
 The results of these papers differ from each other. Our result is close to that of
 Bhattachary, Habib and Mottola (see ref. [3,6]) and 
 differs only by a coefficient: our correction is twice smaller.
\end{abstract}


\maketitle

\section{Introduction}

 An investigation of neutrino oscillations is of great interest for particle physics. 
 Neutrino oscillations in vacuum are described well by the existing theory (see, for example, \cite{Kayser:2008rd}).
 However, the oscillation pattern changes for the propagation in an external gravitational field.
 In the paper we consider neutrino oscillation in the gravitational field described by the Schwarzschild metric.

 The correction to the phase difference of neutrino mass eigenstates due to the gravitational
 field described by the Schwarzschild metric was calculated in papers 
 \cite{ahluwalia-0,bhattacharya-0,ahluwalia-1,ahluwalia-2,bhattacharya-1,fornengo,crocker}. 
 The results obtained in these papers differ from each other.
 In this paper we make our own calculation.

 In the problem of oscillations the spin of a particle is not important,
 so further we consider neutrino as a scalar particle. We also assume
 that the energies of the mass eigenstates which interfere are equal. Arguments in favor of such an assumption
 can be found in papers \cite{stodolsky-0,lipkin,dolgov-0,vysotsky}.

 In Section \ref{schwarzschild} we calculate in the eikonal approximation the phase difference of neutrino mass eigenstates 
 propagating in a radial direction in the gravitational field described by the Schwarzschild metric. 
 In Section \ref{others} we compare our result with results of other authors.

 We use the system of units where $\hbar=c=1$.


\section{Radial propagation in the Schwarzschild metric}
\label{schwarzschild}
 We consider neutrino that was produced at the distance
 $r_A$ from the center of the star which creates the gravitational field and was detected at $r_B$.
 Moreover, the source and the detector are on the line on which the center of the star is located.

 Gravitational field of a spherically symmetric massive body is described by the Schwarzschild metric (see ~\cite{landau}, Eq. (100.14)). 
 In this metric the square of the interval looks like:
 \begin{equation}
  ds^2=g_{\mu\nu}dx^{\mu}dx^{\nu}=\left(1-\frac{r_g}{r}\right)dt^2-\left(1-\frac{r_g}{r}\right)^{-1}dr^2-r^2d\Omega^2,
  \label{schwarzschild_metric}
 \end{equation}
 where $g_{\mu\nu}$ is a metric tensor, $r_g=2GM$ is the Schwarzschild radius ($M$ is the mass of the star),
 $d\Omega$ is the angular part of the metric which is not important for a radial propagation.

 The wave function of neutrino can be written in the form (subscript $a$ corresponds to the $a$-th mass eigenstate):
 \begin{equation}
  \Psi_a({\bf x}, t)=A_a({\bf x}, t)e^{i\Phi_a({\bf x}, t)}.
 \label{wavefunction}
 \end{equation}

 Let us calculate neutrino phase $\Phi_a({\bf x}, t)$ in the eikonal approximation. This approximation
 means that  the variation of $k_{\mu a}=-\partial\Phi_a({\bf x}, t)/\partial x^\mu$ and $A_a({\bf x}, t)$ is small 
 on the wave length. Following \cite{landau}~(\S~53,~\S~101) and \cite{stodolsky-1}~(\S~8) we 
 write the eikonal equation in the form
 \begin{equation}
  g^{\mu\nu}\frac{\partial\Phi_a}{\partial x^\nu}\frac{\partial\Phi_a}{\partial x^\mu}-m_a^2=0,
 \label{eikonaleq}
 \end{equation}
 where $m_a$ is the mass of $a$-th mass eigenstate. 

 The equation (\ref{eikonaleq}) coincides exactly with the Hamilton--Jacobi equation 
 for a classical particle which was solved for the Schwarzschild metric in \cite{landau}~(\S~101). 
 Thus we can use the result for  $\Phi_a({\bf x}, t)$ from \cite{landau}:

\begin{equation}
  \Phi_a=-Et+L\varphi+\int\limits_{r_A}^{r_B}\ls E^2\ls1-\frac{r_g}{r}\rs^{-2}-\ls m_a^2+\frac{L^2}{r^2}\rs\ls1-\frac{r_g}{r}\rs^{-1}\rs^{1/2}dr,
 \label{landau_aswer}
 \end{equation}
 where $E$ is the energy defined at infinity, $L$ is the angular momentum (we consider 
 radial propagation so we put $L=0$).

 Performing Taylor expansion in $m^2/E^2$ up to the second term we get:
 \begin{eqnarray}
 \Phi_a&=&-Et+\int\limits_{r_A}^{r_B}\frac{E}{1-\frac{r_g}{r}}\sqrt{1-\frac{m_a^2}{E^2}\ls1-\frac{r_g}{r}\rs}dr\nonumber\\
 &\approx&-Et+\int\limits_{r_A}^{r_B}\frac{E}{1-\frac{r_g}{r}}\ls1-\frac{m_a^2}{2E^2}\ls1-\frac{r_g}{r}\rs-
 \frac{m_a^4}{8E^4}\ls1-\frac{r_g}{r}\rs^2\rs dr,
 \label{expansion0}
 \end{eqnarray}
 \begin{equation}
  \Phi_a=-Et+\int\limits_{r_A}^{r_B}\ls\frac{E}{1-\frac{r_g}{r}}-\frac{m_a^2}{2E}-
       \frac{m_a^4}{8E^3}\ls1-\frac{r_g}{r}\rs\rs dr.
 \label{expansion}
 \end{equation}

 Then for the phase difference of $a$-th and $a'$-th mass eigenstates we obtain
 \begin{equation}
  \Phi_{aa'}=\Phi_{a'}-\Phi_{a}=\frac{\delta m_{a'a}^2}{2E}\left(r_B-r_A\right)
  +\frac{\delta m_{a'a}^4}{8E^3}\left(r_B-r_A\right)
  -\frac{\delta m_{a'a}^4}{8E^3}r_g\ln \frac{r_B}{r_A},
  \label{our}
 \end{equation}
 where $\delta m_{a'a}^2=m_{a}^2-m_{a'}^2$ and $\delta m_{a'a}^4=m_{a}^4-m_{a'}^4$.

 The first term in the formula (\ref{our}) is the well known expression for the phase difference for 
 the propagation of neutrinos in vacuum. The second term does not depend on $r_g$, so it is not due to 
 a gravitational field. 
 This term is the correction of the order of ${\delta m^4}/{E^3}$ to the first term in the flat space-time.
 Therefore the gravitational correction to the phase difference is described 
 by the third term.
 
 There is a factor of $(1-r_g/r)/8$ in front of ${m_a^4}/{E^3}$ in the formula (\ref{expansion}). 
 It explains why the second and the third terms in the formula (\ref{our}) have the same coefficient $1/8$ and
 opposite signs. Moreover, the coefficients $1/2$ in front of the first term and $1/8$ in front of the second and the third terms
 appear as coefficients of Taylor expansion of $\sqrt{1-x}$ in $x$. 
 Therefore the coefficients in front of the second and the third terms are defined unambiguously
 by the well known coefficient in front of the first term.

 Let us make some numerical estimates of the contribution of the gravitational correction to oscillations of solar neutrinos
 \footnote{In fact, the neutrino oscillation pattern changes because of the Mikheev--Smirnov--Wolfenstein
 effect (see \cite{msw}). But to estimate the order of the gravitational correction
 we do not take this effect into account.}.
 For the Sun we take $r_g=3~\mbox{km}$, $(r_B-r_A)\approx r_B=1.5\cdot10^8~\mbox{km}$, $E=1~\mbox{MeV}$. In order 
 to get the upper limit of the gravitational correction we assume $r_A=r_g$,
 $\delta m^4_{a'a}=2m^2\delta m^2_{a'a}\approx 6.4\cdot10^{-4}~\mbox{eV}^4$ (since 
 $\delta m^4_{a'a}=\delta m^2_{a'a}(m^2_{a}+m^2_{a'})\leq2m^2\delta m^2_{a'a}$),
 where $m=2~\mbox{eV}$ is the present upper bound on the neutrino mass, $\delta m^2_{a'a}=8\cdot10^{-5}~\mbox{eV}^2$
 (see \cite{pdg}, p. 541). Then the phase difference is $\Phi_{aa'}\approx3\cdot10^{7}+6\cdot10^{-5}-2\cdot10^{-11}$
 (each number in this sum equals to the value of the corresponding term in the formula (\ref{our})).
 Since the third term is much smaller then the first and the second terms, the gravitational
 phase shift is not observable.

\section{Comparison with literature}
\label{others}
\subsection{Papers \cite{bhattacharya-0} and~\cite{bhattacharya-1}}
 Our result is close to that obtained in papers~\cite{bhattacharya-0,bhattacharya-1}
 and differs only by a factor in the second and the third
 terms of the formula (\ref{our}). The authors of these papers got $1/4$ while we obtained $1/8$.
 Most likely, there is a mistake in the Taylor expansion of the square root
 in the formula (14) in~\cite{bhattacharya-0} (and in the formula (12) in \cite{bhattacharya-1}).
 In a conversation with us during the 4-th Sakharov Conference on May 22, 2009 
 one of these authors, Emil Mottola, said that he admitted the possibility of
 such an error.

\subsection{Papers \cite{fornengo} and~\cite{crocker}}
 In the papers \cite{fornengo,crocker} the gravitational correction
 was calculated only in the order $\delta m^2/E$. The following expression
 for the phase difference of neutrino mass eigenstates was obtained:
\begin{equation}
 \Phi_{aa'}=\frac{\delta m^2_{a'a}}{2E}\left|r_B-r_A\right|.
\label{34I}
\end{equation}
 
 The absence of the influence of gravity in the first order in $\delta m^2/E$ is
 in agreement with our result (see the formula (\ref{our})). The next order in $\delta m^2/E$
 where gravity appears non-trivially was not considered in \cite{fornengo,crocker}.

 But having obtained the absence of the influence of gravity in the first order in $\delta m^2/E$
 the authors say that gravity appears implicitly. 
 To show this they rewrite global variables in terms of local variables.

 As local variables the authors of \cite{fornengo,crocker} use local energy $E^{({\rm loc})}(r_B)={E}\left(1-{r_g}/{r_B}\right)^{-1/2}$
 and proper distance $L_p$ between the points $r_A$ and $r_B$: 
 \begin{equation}
  L_p=\int\limits_{r_A}^{r_B}\sqrt{g_{rr}}dr\approx\int\limits_{r_A}^{r_B}\ls1+\frac{r_g}{2r}\rs dr=
  (r_B-r_A)+\frac{r_g}{2}\ln\frac{r_B}{r_A}.
  \nonumber
  \label{38I}
 \end{equation}
Here they keep only the first order in $r_g$ in the second equality.

Then the formula (\ref{34I}) can be rewritten as follows
\begin{eqnarray}
 \Phi_{aa'}=\frac{\delta m^2_{a'a}}{2E}\left|r_B-r_A\right|&\approx&\frac{\delta m^2_{a'a}L_p}{2E^{({\rm loc})}(r_B)}
              \frac{\ls 1-\frac{r_g}{2L_p}\ln\frac{r_B}{r_A}\rs}{\sqrt{1-\frac{r_g}{r_B}}}\nonumber\\
	   &\approx&\frac{\delta m^2_{a'a}L_p}{2E^{({\rm loc})}(r_B)}\ls1-\frac{r_g}{2L_p}\ls\ln\frac{r_B}{r_A}-\frac{L_p}{r_B}\rs\rs\nonumber.
\end{eqnarray}

Such a transformation from global to local variables can be done for any quantity. 
Therefore, the obtained apparent dependence on gravity is not specific for the physics of neutrino 
oscillations. 

\subsection{Papers \cite{ahluwalia-0},~\cite{ahluwalia-1} and~\cite{ahluwalia-2}}
 The result for the gravitational correction obtained in the papers
 \cite{ahluwalia-0,ahluwalia-1,ahluwalia-2} 
 differs greatly from ours.
 The authors of these papers state that the nontrivial correction arises already 
 in the terms of the order of $\delta m^2/E$.


\acknowledgments
 
 We are thankful to L.~B. Okun, M.~I. Vysotsky, A.~D. Dolgov, M.~B. Voloshin and M.~V. Rotaev for valuable remarks and discussions.

 This work was supported by the RFBR grants 07-02-00830-a, 08-02-00494-a and by the grant SH-4568.2008.2.


\begin{thebibliography}{16}
\expandafter\ifx\csname natexlab\endcsname\relax\def\natexlab#1{#1}\fi
\expandafter\ifx\csname bibnamefont\endcsname\relax
  \def\bibnamefont#1{#1}\fi
\expandafter\ifx\csname bibfnamefont\endcsname\relax
  \def\bibfnamefont#1{#1}\fi
\expandafter\ifx\csname citenamefont\endcsname\relax
  \def\citenamefont#1{#1}\fi
\expandafter\ifx\csname url\endcsname\relax
  \def\url#1{\texttt{#1}}\fi
\expandafter\ifx\csname urlprefix\endcsname\relax\def\urlprefix{URL }\fi
\providecommand{\bibinfo}[2]{#2}
\providecommand{\eprint}[2][]{\url{#2}}

\bibitem[{\citenamefont{Kayser}(2008)}]{Kayser:2008rd}

\refitem{article}
\bibinfo{author}{\bibfnamefont{B.}~\bibnamefont{Kayser}},
  \bibinfo{journal}{Phys. Lett. B} \textbf{\bibinfo{volume}{667}},
  \bibinfo{pages}{163} (\bibinfo{year}{2008}), \eprint{hep-ph/0804.1497}.

\bibitem[{\citenamefont{Ahluwalia and Burgard}(1996)}]{ahluwalia-0}

\refitem{article}
\bibinfo{author}{\bibfnamefont{D.~V.} \bibnamefont{Ahluwalia}}
  \bibnamefont{and} \bibinfo{author}{\bibfnamefont{C.}~\bibnamefont{Burgard}},
  \bibinfo{journal}{Gen. Rel. Grav.} \textbf{\bibinfo{volume}{28}},
  \bibinfo{pages}{1161} (\bibinfo{year}{1996}), \eprint{gr-qc/9603008}.

\bibitem[{\citenamefont{Bhattacharya
  \emph{et~al.}}()\citenamefont{Bhattacharya, Habib, and
  Mottola}}]{bhattacharya-0}

\refitem{misc}
\bibinfo{author}{\bibfnamefont{T.}~\bibnamefont{Bhattacharya}},
  \bibinfo{author}{\bibfnamefont{S.}~\bibnamefont{Habib}}, \bibnamefont{and}
  \bibinfo{author}{\bibfnamefont{E.}~\bibnamefont{Mottola}},
  \eprint{gr-qc/9605074}.

\bibitem[{\citenamefont{Ahluwalia and Burgard}()}]{ahluwalia-1}

\refitem{misc}
\bibinfo{author}{\bibfnamefont{D.~V.} \bibnamefont{Ahluwalia}}
  \bibnamefont{and} \bibinfo{author}{\bibfnamefont{C.}~\bibnamefont{Burgard}},
  \eprint{gr-qc/9606031}.

\bibitem[{\citenamefont{Ahluwalia and Burgard}(1998)}]{ahluwalia-2}

\refitem{article}
\bibinfo{author}{\bibfnamefont{D.~V.} \bibnamefont{Ahluwalia}}
  \bibnamefont{and} \bibinfo{author}{\bibfnamefont{C.}~\bibnamefont{Burgard}},
  \bibinfo{journal}{Phys. Rev. D} \textbf{\bibinfo{volume}{57}},
  \bibinfo{pages}{4724} (\bibinfo{year}{1998}), \eprint{gr-qc/9803013}.

\bibitem[{\citenamefont{Bhattacharya
  \emph{et~al.}}(1999)\citenamefont{Bhattacharya, Habib, and
  Mottola}}]{bhattacharya-1}

\refitem{article}
\bibinfo{author}{\bibfnamefont{T.}~\bibnamefont{Bhattacharya}},
  \bibinfo{author}{\bibfnamefont{S.}~\bibnamefont{Habib}}, \bibnamefont{and}
  \bibinfo{author}{\bibfnamefont{E.}~\bibnamefont{Mottola}},
  \bibinfo{journal}{Phys. Rev. D} \textbf{\bibinfo{volume}{59}},
  \bibinfo{pages}{067301} (\bibinfo{year}{1999}).

\bibitem[{\citenamefont{Fornengo \emph{et~al.}}(1997)\citenamefont{Fornengo,
  Giunti, Kim, and Song}}]{fornengo}

\refitem{article}
\bibinfo{author}{\bibfnamefont{N.}~\bibnamefont{Fornengo}},
  \bibinfo{author}{\bibfnamefont{C.}~\bibnamefont{Giunti}},
  \bibinfo{author}{\bibfnamefont{C.~W.} \bibnamefont{Kim}}, \bibnamefont{and}
  \bibinfo{author}{\bibfnamefont{J.}~\bibnamefont{Song}},
  \bibinfo{journal}{Phys. Rev. D} \textbf{\bibinfo{volume}{56}},
  \bibinfo{pages}{1895} (\bibinfo{year}{1997}), \eprint{hep-ph/9611231}.

\bibitem[{\citenamefont{Crocker \emph{et~al.}}(2004)\citenamefont{Crocker,
  Giunti, and Mortlock}}]{crocker}

\refitem{article}
\bibinfo{author}{\bibfnamefont{R.~M.} \bibnamefont{Crocker}},
  \bibinfo{author}{\bibfnamefont{C.}~\bibnamefont{Giunti}}, \bibnamefont{and}
  \bibinfo{author}{\bibfnamefont{D.~J.} \bibnamefont{Mortlock}},
  \bibinfo{journal}{Phys. Rev. D} \textbf{\bibinfo{volume}{69}},
  \bibinfo{pages}{063008} (\bibinfo{year}{2004}), \eprint{hep-ph/0308168}.

\bibitem[{\citenamefont{Stodolsky}(1998)}]{stodolsky-0}

\refitem{article}
\bibinfo{author}{\bibfnamefont{L.}~\bibnamefont{Stodolsky}},
  \bibinfo{journal}{Phys. Rev. D} \textbf{\bibinfo{volume}{58}},
  \bibinfo{pages}{036006} (\bibinfo{year}{1998}), \eprint{hep-ph/9802387}.

\bibitem[{\citenamefont{Lipkin}()}]{lipkin}

\refitem{misc}
\bibinfo{author}{\bibfnamefont{H.~J.} \bibnamefont{Lipkin}},
  \eprint{hep-ph/0212093}.

\bibitem[{\citenamefont{Dolgov \emph{et~al.}}(2005)}]{dolgov-0}

\refitem{article}
\bibinfo{author}{\bibfnamefont{A.~D.} \bibnamefont{Dolgov}}
  \bibnamefont{\emph{et~al.}}, \bibinfo{journal}{Nucl. Phys. B}
  \textbf{\bibinfo{volume}{729}}, \bibinfo{pages}{79} (\bibinfo{year}{2005}),
  \eprint{hep-ph/0505251}.

\bibitem[{\citenamefont{Vysotsky}(2003)}]{vysotsky}

\refitem{article}
\bibinfo{author}{\bibfnamefont{M.}~\bibnamefont{Vysotsky}},
  \bibinfo{journal}{Surveys High Energ. Phys.} \textbf{\bibinfo{volume}{18}},
  \bibinfo{pages}{19} (\bibinfo{year}{2003}), \eprint{hep-ph/0307218}.

\bibitem[{\citenamefont{{L.~D.~Landau,~E.~M.~Lifshitz}}(1986)}]{landau}

\refitem{book}
\bibinfo{author}{\bibnamefont{{L.~D.~Landau,~E.~M.~Lifshitz}}},
  \emph{\bibinfo{title}{Classical Theory of Fields}}
  (\bibinfo{publisher}{Fizmatlit, Moscow}, \bibinfo{year}{1986}).

\bibitem[{\citenamefont{Stodolsky}(1979)}]{stodolsky-1}

\refitem{article}
\bibinfo{author}{\bibfnamefont{L.}~\bibnamefont{Stodolsky}},
  \bibinfo{journal}{Gen. Rel. Grav.} \textbf{\bibinfo{volume}{11}},
  \bibinfo{pages}{391} (\bibinfo{year}{1979}).

\bibitem[{\citenamefont{Mikheev}(1987)}]{msw}

\refitem{article}
\bibinfo{author}{\bibfnamefont{A.~Y.} \bibnamefont{Mikheev},
  \bibfnamefont{S.~P.~Smirnov}}, \bibinfo{journal}{Sov. Phys. Usp.}
  \textbf{\bibinfo{volume}{30}}, \bibinfo{pages}{759} (\bibinfo{year}{1987}).

\bibitem[{\citenamefont{Amsler \emph{et~al.}}(2008)}]{pdg}

\refitem{article}
\bibinfo{author}{\bibfnamefont{C.}~\bibnamefont{Amsler}}
  \bibnamefont{\emph{et~al.}} (\bibinfo{collaboration}{Particle Data Group}),
  \bibinfo{journal}{Phys. Lett. B} \textbf{\bibinfo{volume}{667}},
  \bibinfo{pages}{1} (\bibinfo{year}{2008}).

\end{thebibliography}

\end{document}